\documentclass[a4paper,12pt]{scrartcl}
\usepackage[utf8]{inputenc}
\usepackage[T1]{fontenc}
\usepackage[english]{babel}
\usepackage{csquotes}

\usepackage[section]{placeins}
\usepackage{graphicx}
\usepackage{caption}
\usepackage{subcaption}
\usepackage{varioref}
\usepackage[breaklinks=true,colorlinks=false, pdfborder={0 0 0}]{hyperref}

\usepackage{siunitx}

\usepackage[shortcuts]{extdash}

\usepackage{amsmath}
\usepackage{amsfonts}
\usepackage{amssymb}

\usepackage[backend=biber,style=numeric,autopunct, sorting=none, alldates=year, block=ragged, autolang=hyphen, isbn=false]{biblatex}
\usepackage[nottoc]{tocbibind}

\graphicspath{ {.} }
\DeclareGraphicsExtensions{.pdf,.png,.jpg}

\title{ROPPERI \\-\\ A TPC readout with GEMs, pads and Timepix}
\author{Ulrich Einhaus\thanks{DESY Hamburg}, Jochen Kaminksi\thanks{University of Bonn}}
\date{\large Talk presented at the International Workshop on Future Linear Colliders (LCWS2016), Morioka, Japan, 5-9 December 2016. C16-12-05.4.}

\begin{document}

\maketitle

\begin{abstract}
The concept of a hybrid readout of a time projection chamber is presented. It combines a GEM-based amplification and a pad-based anode plane with a pixel chip as readout electronics. This way, a high granularity enabling to identify electron clusters from the primary ionisation is achieved as well as flexibility and large anode coverage. The benefits of this high granularity, in particular for dE/dx measurements, are outlined and the current software and hardware development status towards a proof-of-principle is given.
\end{abstract}

\section{Introduction}
\label{sec:introduction}

\begin{figure}[thp]
  \centering
  \includegraphics[width=\textwidth,height=0.35\textheight,keepaspectratio=true]{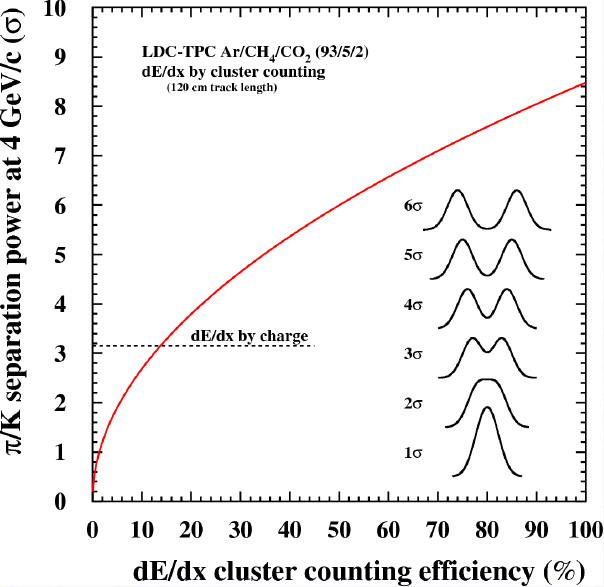}
  \caption{Pion/kaon separation power of dE/dx by cluster counting, from \cite{Hauschild06}.}
  \label{fig:piokaonsep}
\end{figure}

For the International Large Detector (ILD) \cite{ILD_TDR_detectors} at the International Linear Collider (ILC) \cite{ILCSummary} a Time Projection Chamber (TPC) \cite{TPC_CC} is foreseen as central tracker. A key feature of this gaseous detector is the inherent dE/dx capability: Since the relation of momentum and energy loss (dE/dx) of a traversing particle depends on its rest mass, and thus its species, measuring both properties allows for a particle identification determined from the Bethe-Bloch-curve. The energy loss is conventionally measured by summing all electrons generated from the ionisation by the incident particle. For each ionising interaction the number of generated electrons is given by a Landau distribution which has a long tail towards large numbers of electrons. The relatively large width of this distribution worsens the correlation of the measured energy and the momentum of the particle. It is advantageous to instead count the number of ionising interactions the incident particle goes through. This is given by a Poissonian distribution with a significantly smaller width, resulting in a better correlation and particle identification power, as demonstrated in \cite{Hauschild06}. In \autoref{fig:piokaonsep}, the separation power for pion/kaon-separation depending on the cluster counting efficiency is shown compared to the conventional dE/dx by charge summation. In former experiments with prototypes, a cluster counting efficiency of only \SI{20}{\percent}-\SI{30}{\percent} was reached. Nevertheless, the resulting separation power is still better than by charge summation. Also, improved algorithms are expected to deliver a higher cluster counting efficiency. However, cluster counting can only work if a sufficient correlation between the position of the electrons of one cluster is preserved during drift. This needs to be investigated in simulation.

\begin{figure}[thp]
   \centering
   \includegraphics[width=0.95\textwidth,height=0.3\textheight,scale=1,keepaspectratio=true]{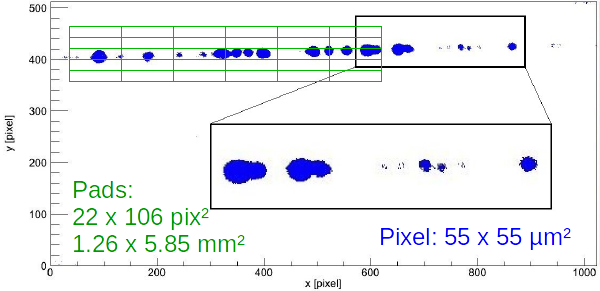}
   \caption{Event display of a track recorded with a Timepix Octoboard, from \cite{Lupberger14:1}. The activated pixels are shown in blue, the green overlay shows the pitch of a typical pad-based readout.}
   \label{fig:event_display}
 \end{figure}

To achieve the cluster counting capability, a sufficiently high granularity is needed for the TPC readout system. A comparison of two existing prototype systems is shown in \autoref{fig:event_display}. The electron clusters are amplified by GEMs (Gas Electron Multipliers) \cite{SAULI1997531}, creating charge clouds visible as blue blobs. The anode consists of 8 Timepix ASICs with their pixels as immediate sensitive anode, which thus has a pitch of \SI{55x55}{} \si{\micro\meter ^2}. The charge clouds are clearly identifiable – the granularity is even higher than needed, causing more data than necessarily required. On the other hand, the green overlay of a typical current pad-based readout system (GridGEM \cite{lctpc16}), shows, that with its pads of ca. \SI{1x6}{} \si{mm^2} a cluster identification is not possible. Therefore, we propose a novel readout structure, called ROPPERI (Readout Of a Pad Plane with ElectRonics designed for pIxels), that allows pad sizes of around \SI{300}{\micro\meter} to enable cluster counting, gives a large flexibility and keeps the channel number low at the same time.

\section{Readout Board Structure}
\label{sec:board}

\begin{figure}[thp]
  \centering
  \includegraphics[width=0.95\textwidth,height=0.25\textheight,scale=1,keepaspectratio=true]{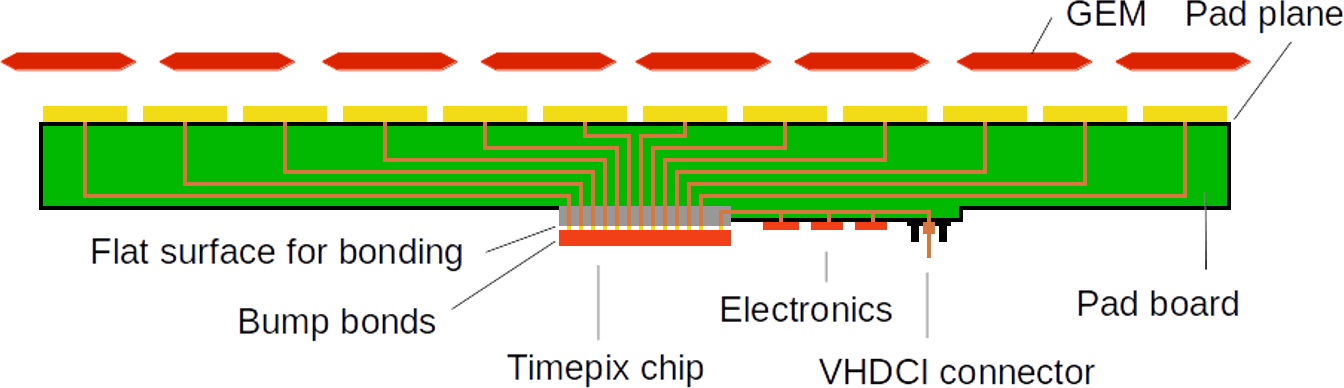}
  \caption{ROPPERI structure schematic.}
  \label{fig:routing_structure}
\end{figure}

\begin{figure}[thp]
  \centering
  \includegraphics[width=0.95\textwidth,height=0.4\textheight,scale=1,keepaspectratio=true]{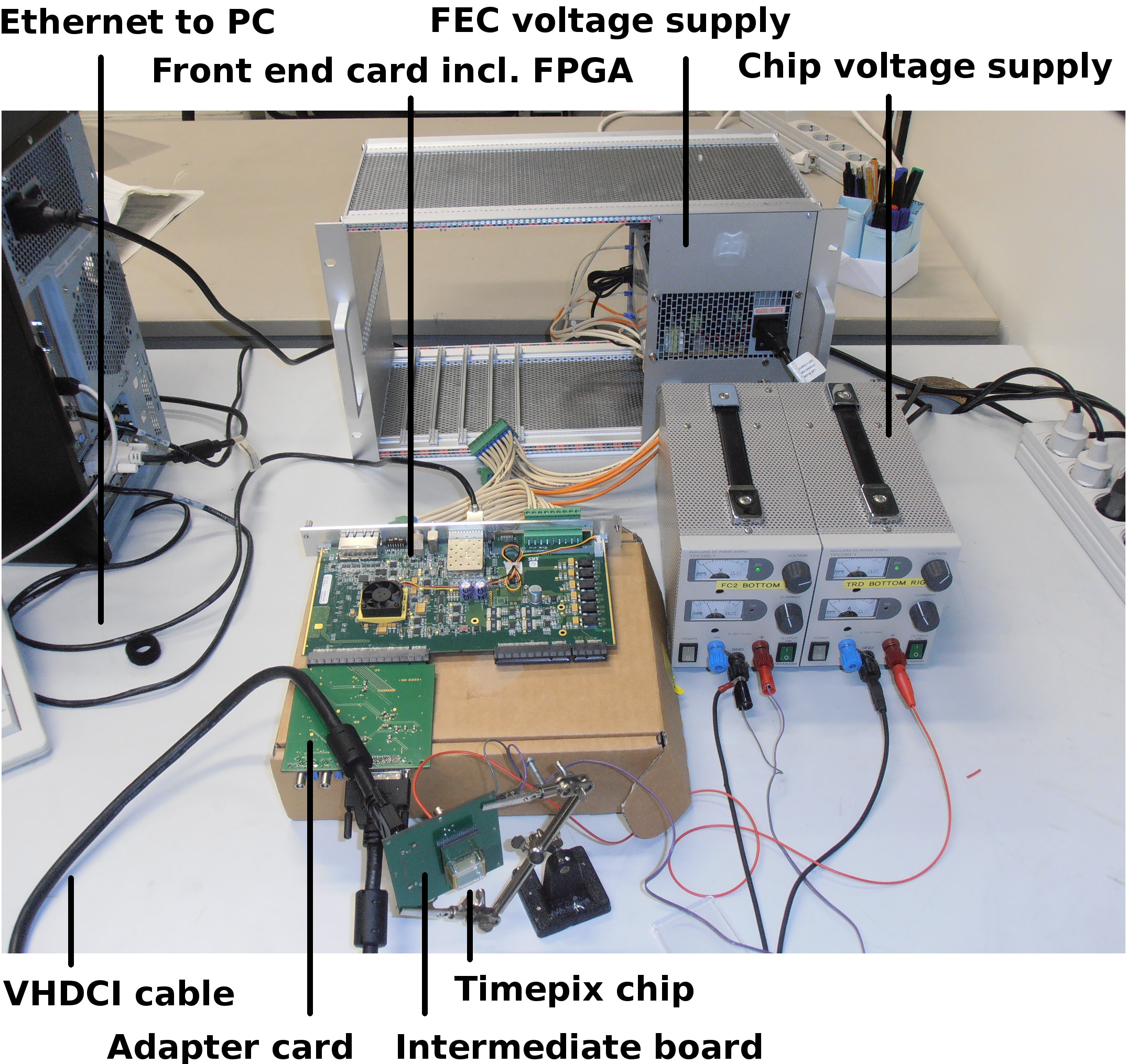}
  \caption{Timepix readout setup.}
  \label{fig:full_setup}
\end{figure}

The ansatz for a ROPPERI board implementation is shown in \autoref{fig:routing_structure}: GEMs are used for amplification, small pads on a PCB form the anode and are read out by a Timepix ASIC, which has a matrix of \SI{256x256}{} $=65,536$ pixels. The connections from the pads are routed through the PCB to the ASIC which is bump bonded to the PCB surface, that needs to be sufficiently flat. Technology-wise, it is pioneering work to connect a pixel chip with such a small pitch directly to a PCB. The Timepix power and communication pads are on the same side of the ASIC as the pixels. They are usually connected by wire bonds. In the ROPPERI approach also these pads have to be connected by bump bonds to the PCB, which also hosts the further electronics elements including the connectors for the chip voltage supply and an I/O-cable plug. The data processing is conducted by the SRS (Scalable Readout System) developed by the RD51 group at CERN. An FEC (Front-End Concentrator card) hosts an FPGA (Field Programmable Gate Array) that reads the data from the Timepix ASIC via an adapter card and a VHDCI-cable to the ROPPERI board, for a detailed implementation see \cite{Lupberger16}. The FEC can be connected via Ethernet to a PC, see \autoref{fig:full_setup}

\subsection{Benefits Compared to Existing Systems}
\label{sec:benefits}

Compared to the pad-based systems, like the GridGEM \cite{lctpc16}, ROPPERI can have a higher granularity leading to an improved double hit/track resolution, to a reduced occupancy and in particular to the charge cloud and possibly cluster identification capability. In addition, the Timepix ASIC allows for a significantly smaller footprint of the readout electronics.\\
Compared to the pixel-based systems, in particular the InGrid system \cite{Lupberger14:2}, ROPPERI has a lower granularity, but measures each primary electron with several pads, allowing for a position calculation using a fit. Instead, the InGrid system records each electron with exactly one pixel. Pixel granularity combined with a GEM, as in \cite{Renz09}, does not increase the performance, but only the data output. In addition, the pixel-based anode can currently only cover ca. \SI{50}{\percent} of the anode area in a setup with 12 octoboards \cite{Lupberger11}, or up to \SI{63}{\percent} with tighly stacked ASICs as suggested in \cite{Timmermans11}, because the ASICs needs some overhead area, and the module geometry of the foreseen ILD TPC is not square, like the ASICs. ROPPERI's separate anode PCB allows for a coverage of more than \SI{90}{\percent}, comparable to the traditional pad-based systems. Also, it is more flexible with regard to the granularity, since for a desired change in pad size it is required to only produce new PCBs and bond them to the ASICs, not to produce new ASICs with a different pitch.

\section{Hardware}
\label{sec:hardware}

\subsection{Current Status}
\label{sec:hardware_status}

\begin{figure}[thp]
  \centering
  \includegraphics[width=0.95\textwidth,height=0.4\textheight,scale=1,keepaspectratio=true]{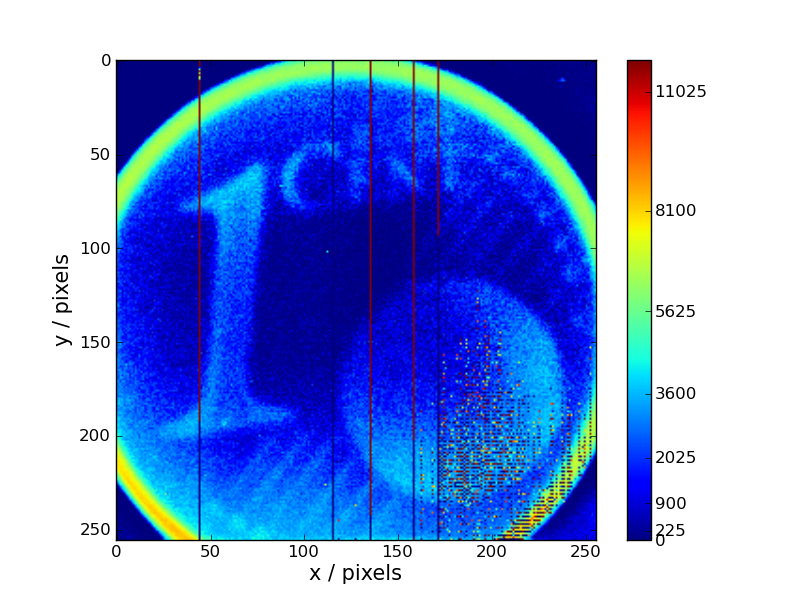}
  \caption{Display of data taken from a Timepix ASIC which had an electrically pulsed coin capacitatively coupled to it. Some columns are damaged and on the bottom right there are some very noisy pixels.}
  \label{fig:TP_cent}
\end{figure}

\begin{figure}[thp]
  \centering
  \includegraphics[width=\textwidth,height=0.75\textheight,keepaspectratio=true]{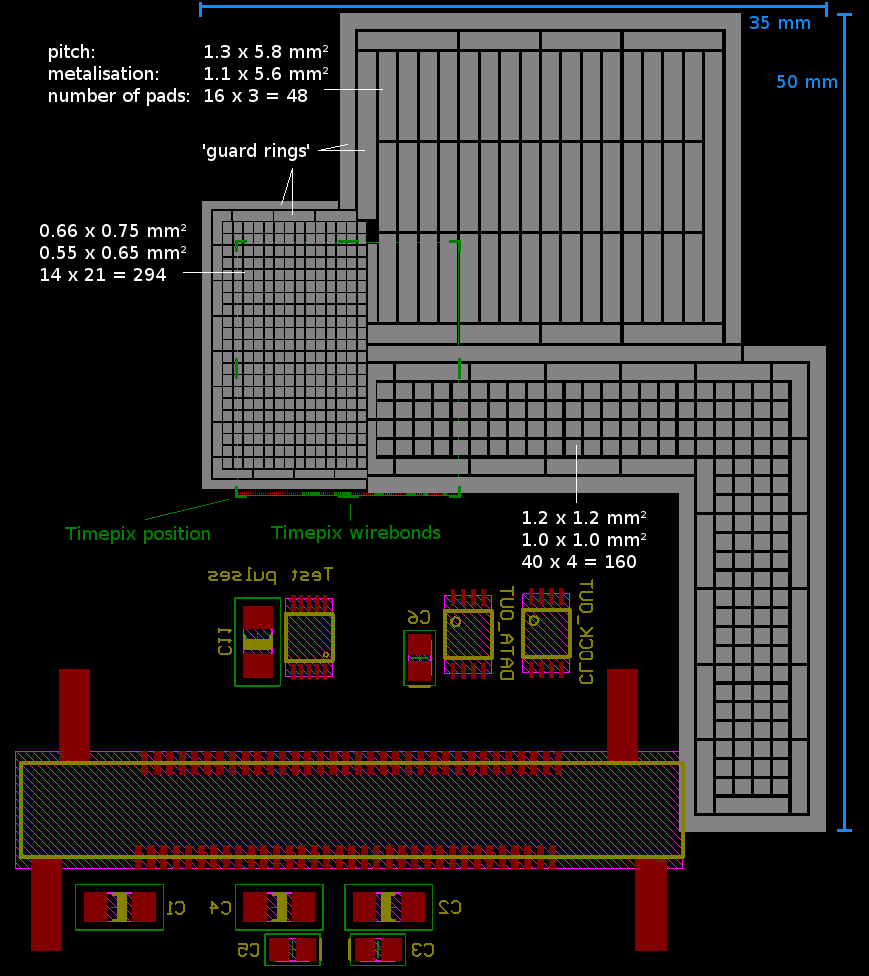}
  \caption{Padlayout of the first ROPPERI prototype board.}
  \label{fig:padlayout}
\end{figure}

The SRS has been set up and successfully tested. For this, a single Timepix ASIC was mounted close to an electrically pulsed coin and read out. Through the capacitative coupling a signal was induced to the pixels. The resulting ADC data map of the chip shows the coin, see \autoref{fig:TP_cent}. The dead columns and noisy pixels are part of the low-quality chip used for the tests. For the actual test system, high quality chips are available with a significantly lower count of dead or noisy channels.
The board for the test system has been designed and ordered, and is expected to be delivered before the end of 2016. \autoref{fig:padlayout} shows the main features of the first board: Three different pad sizes are used for the sensitive pads, shown in grey. The currently minimal achievable pad pitch is \SI{0.66x0.75}{} \si{mm^2}, used for about 300 of the pads placed directly above the Timepix ASIC (envelope shown in green). The \SI{1.2x1.2}{} \si{mm^2} pads are used to check signal quality for different line lengths. An array of \SI{1.3x5.8}{} \si{mm^2} pads corresponds to the standard pad pitch of FADC systems for comparison. The sensitive pads are surrounded by grounded 'guard ring' pads to reduce cross talk. In total 502 pads are connected to the chip and cover a sensitive area of a few \si{cm^2}. In the lower part of \autoref{fig:padlayout} the placements of some passive elements and the VHDCI connector are shown in red. The board is going to be used with a triple GEM stack of standard CERN \SI{10x10}{} \si{cm^2} GEMs in a modular prototype TPC. This prototype has a maximum drift length of 5 \si{cm} and can be used with an integrated radioactive source.

\subsection{Challenges}
\label{sec:hardware_challenges}

\begin{figure}[thp]
  \centering
  \includegraphics[width=\textwidth,height=0.5\textheight,keepaspectratio=true]{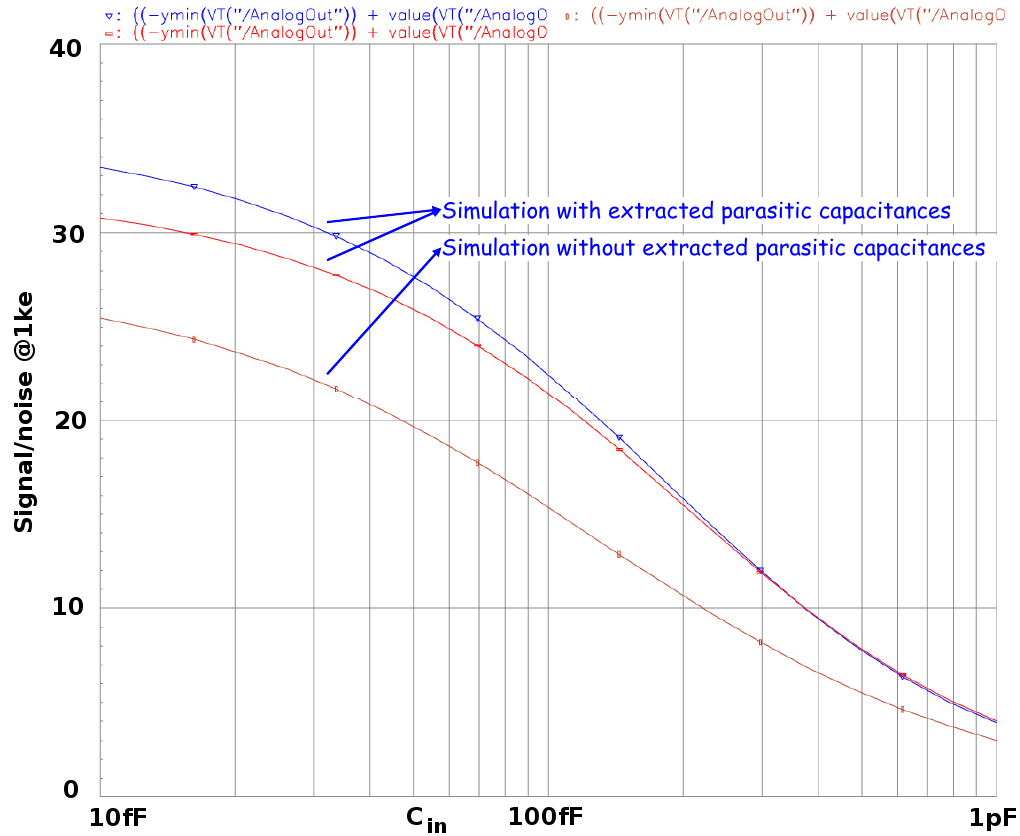}
  \caption{Simulated signal-to-noise ratio of the Timepix ASIC depending on the attached input capacitance, from \cite{Llopart06}.}
  \label{fig:capacitance}
\end{figure}

The discrepancy between the pad pitch of ca. \SI{700}{\micro\meter} and a desired pitch of \SI{300}{\micro\meter} or less results from technical limitations of the routing in the board. The first test board is made from standard FR-4 material, which allows for a minimum feature size of \SI{80}{\micro\meter} and a minimum via size of \SI{300}{\micro\meter} or larger. Further requirements on buried vias and air tightness in the end result in the given limit on the pad pitch. For future boards, other base materials will be taken into account, such as ceramic which has a feature size around \SI{10}{\micro\meter} and a minimum via size of \SI{100}{\micro\meter}. This still does not allow to use the full Timepix ASIC with its \SI{55}{\micro\meter} pitch, but fulfills the cluster finding requirements.
Another challenge is the input capacitance of the pixels: Usually, pixel chips are bonded to sensors with the same pitch and small conductor volume and thus a small capacitance per sensor channel. Therefore, the Timepix ASIC is made for input capacitances below \SI{100}{fF}, compare \autoref{fig:capacitance} \cite{Llopart06}. In case of the ROPPERI system, each connection goes through the board, and the capacitance is dominated by the line length with a value around 1 pF / 25 mm. The pads and the bump bonds add only around \SI{100}{fF} to the input capacitance. For line lengths between pad and pixel in the order of cm this clearly reduces the expected signal to noise ratio to a difficult level. This can probably be mitigated by an increased gas amplicifation in the GEM stack, leading to a signal of up to several 10k electrons. In total, a solution to the capacitance challenge seems feasible and will be investigated with the first test board. In particular, the smallest pads are placed directly above the chip to have a capacitance as small as possible, and the medium size pads are designed to have different line lengths to investigate the length dependence of the signal-to-noise ratio.

\subsection{Next Steps}
\label{sec:hardware_next_steps}

\begin{figure}[thp]
  \centering
  \includegraphics[width=\textwidth,height=0.5\textheight,keepaspectratio=true]{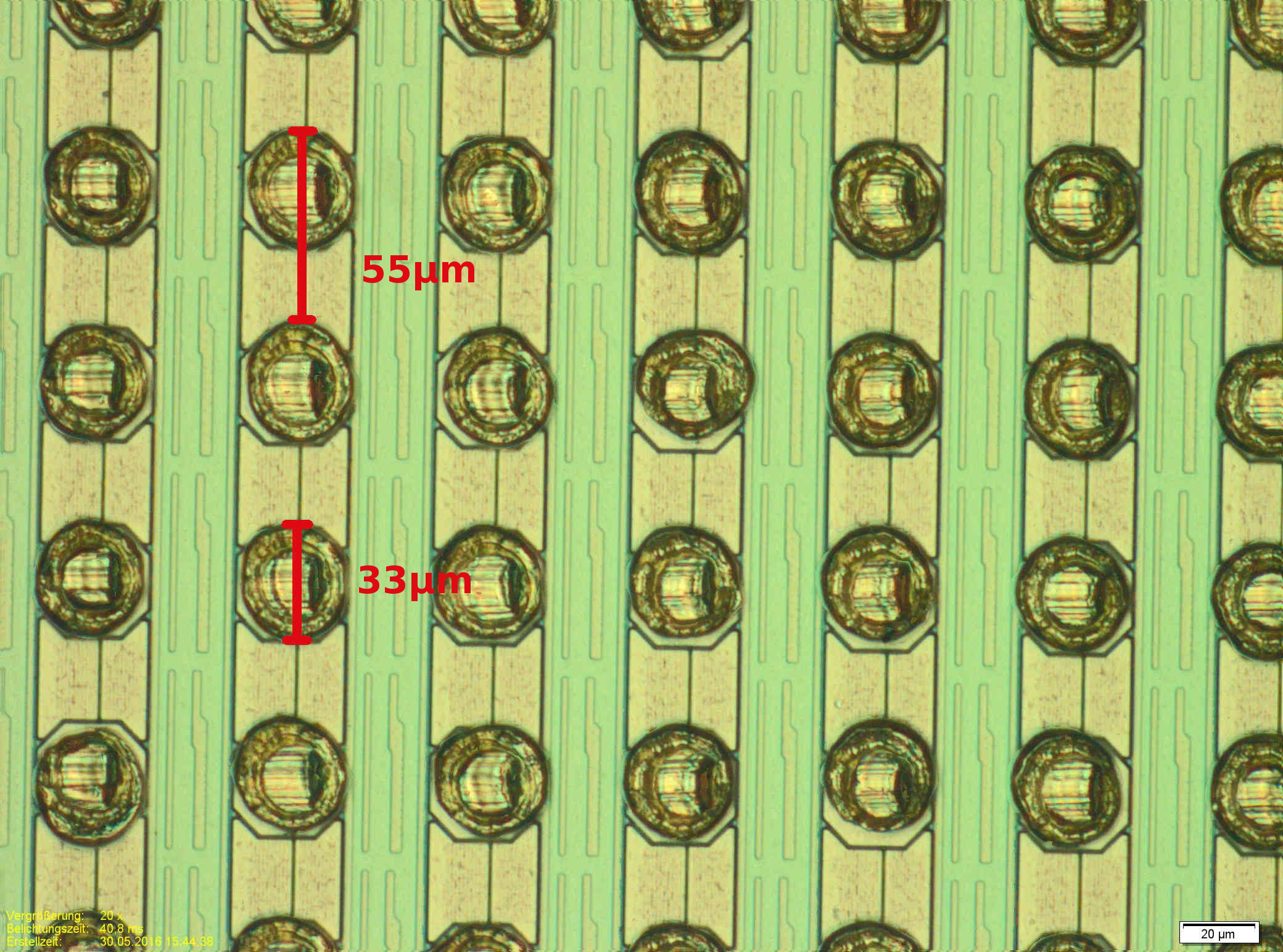}
  \caption{Microscope picture of the Timepix surface after gold stud application.}
  \label{fig:bonding}
\end{figure}

The next step after production of the board is bonding the Timepix ASIC to it. A collaboration with the KIT in Karlsruhe has been established. Their lab has the equipment to conduct gold stud bonding at the required level. This means application of gold studs to each of the chip's pixels and the board's bonding pads, as well as flip chip bonding the two components. A test of the application of gold studs to the Timepix surface was successfully performed in mid 2016, see \autoref{fig:bonding}. The Timepix pitch made it necessary to change the gold wire and its capillary to \SI{15}{\micro\meter}. Application of gold studs to the board will be tested after their delivery. The boards have an 'ENEPIG' surface coating known to be well suited for bonding.
The equipping with the passive elements and the VHDCI plug will be done manually not to risk damaging the potentially fragile bonding connection between chip and PCB. After this, the board will go through a principle working test, in particular with respect to the signal-to-noise ratio. More detailed analyses will follow, such as signal-to-noise depending on pad size and internal line length, and track reconstruction. Furthermore, the development of a more sophisticated board is envisaged, using a different base material like ceramics to allow for sufficiently small pad sizes to study cluster identification and achieve a full coverage of the anode area. A long-term goal is the construction of a dedicated readout module to be used in the Large Prototype TPC (LPTPC) \cite{Lentdecker09} developed by the Linear Collider TPC collaboration (LCTPC).

\section{Simulation}
\label{sec:simulation}

\begin{figure}[thp]
  \centering
  \includegraphics[width=\textwidth,height=0.5\textheight,keepaspectratio=true]{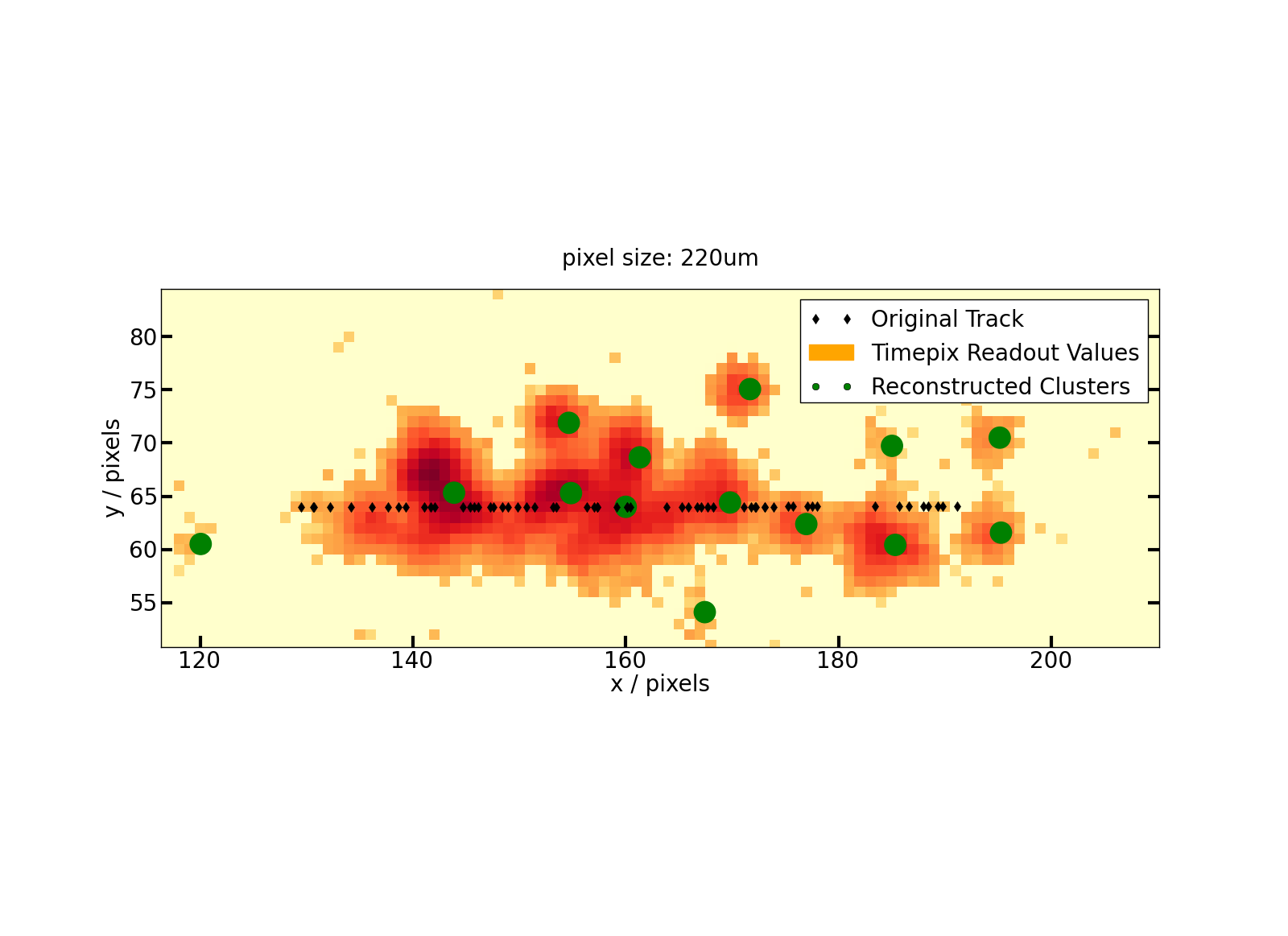}
  \caption{Event display of a simulated track, the resulting Timepix readout and the hits after reconstruction with the Source Extractor software.}
  \label{fig:source_extractor}
\end{figure}

\begin{figure}[thp]
  \centering
  \includegraphics[width=\textwidth,height=0.35\textheight,keepaspectratio=true]{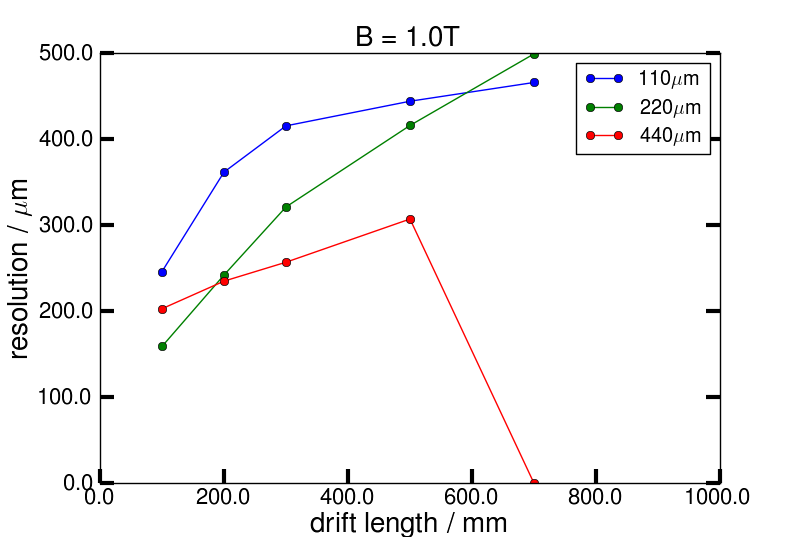}
  \caption{Point resolution plot as result of the full ROPPERI simulation chain. Work in progress!}
  \label{fig:point_resolution}
\end{figure}

The prototype system development is accompanied by intense simulation studies to determine the optimal setup of the ROPPERI board, in particular regarding pad size and cluster identification. For this, a simulation chain has been implemented within the MarlinTPC software framework. It includes the production of MC-particles in the detector volume, the drift, amplification and registration of the electrons on the ROPPERI pad plane as well as digitising the signal including noise. Then a reconstruction algorithm based on the 'Source Extractor' software package follows. This is an external software package originating from astrophysics to scan sky maps for sources and reconstruct them. The algorithm has been adapted to scan data from the Timepix readout for charge clouds. The necessary data conversion to the .fits-format, the execution of the Source Extractor and the re-import to .slcio were been implemented in the scope of a bachelor thesis \cite{Deisting12} by Alexander Deisting at the University of Bonn. The resulting reconstructed hits from the charge clouds are then used as input for a track reconstruction. The track can be subsequently analised, e. g. regarding the point resolution. The whole simulation and reconstruction chain has been implemented and first sets of Monte Carlo data have been produced. An example of the Source Extractor output is shown in \autoref{fig:source_extractor}, and in \autoref{fig:point_resolution}, a resulting plot for the point resolution is shown. This work is still ongoing. Especially the reconstruction chain needs to be further tuned to process the data in a more efficient and reliable way.

\section{Conclusion}
\label{sec:conclusion}

The ROPPERI concept allows for a TPC readout with a high level of integration and a high granularity at the same time. The foreseen granularity of a few hundred \si{\micro\meter} is expected to not only provide lower occupancy and better double-track separation, but to also significantly improve the TPC-intrinsic dE/dx capability via cluster counting. At the same time, the advantages of a pad-based readout are preserved, in particular its flexibility. 
The current hardware development was introduced in detail. It will result in a test of the first prototype within the next months, which will hopefully deliver a proof-of-principle of this technology. In parallel, a software framework has been set up to investigate the future prospects and study more detailed properties using Monte Carlo simulations.

\section{Acknowledgements}
\label{sec:acknowledgements}

We would like to thank Michele Caselle and Fabio Colombo from the Karlsruhe Institute for Technology (KIT) for their efforts regarding the gold stud bonding of the Timepix ASIC.
We thank our DESY colleagues for enduring support, in particular Oliver Schäfer who built the UNIMOCS TPC that will be used for the first tests.
Thanks also to DESY for funding and technical support.


\printbibliography
\end{document}